\documentclass[11pt,showpacs,preprintnumbers,floatfix]{article}
\pdfoutput=1
\usepackage{graphicx} 
\usepackage[colorlinks=true,linkcolor=blue,citecolor=blue,urlcolor=blue,linktocpage=true]{hyperref}
\usepackage{dcolumn} 
\usepackage{bm} 
\usepackage{amsfonts,amsthm,amsmath,amssymb}
\usepackage{soul}
\usepackage{epsfig}
\usepackage{xcolor}
\usepackage{textcomp}
\usepackage{hyperref}
\usepackage{titlesec}
\usepackage{slashed}
\usepackage{caption}
\usepackage{subcaption}
\usepackage{siunitx}
\usepackage{booktabs}
\usepackage{multirow}
\usepackage{cite}
\usepackage{placeins}
\usepackage{comment}
\usepackage{braket}
\usepackage{latexsym}
\def\be{\begin{equation}}
\def\ee{\end{equation}}
\def\bea{\begin{eqnarray}}
\def\eea{\end{eqnarray}}

\def\bdm{\begin{displaymath}} 
\def\edm{\end{displaymath}}

\def\bq{\begin{quote}}
\def\eq{\end{quote}}

 at 10truept

\newcommand{\bi}{\begin{itemize}}
\newcommand{\ei}{\end{itemize}}

\newcommand{\beq}{\begin{equation}}
\newcommand{\eeq}{\end{equation}}
\newcommand{\p}{\partial}
\newcommand{\ep}{\epsilon}
\newcommand{\lle}{\left<}
\newcommand{\rgr}{\right>}

\newcommand{\vect}[1]{\bm{\mathrm{{#1}}}}

\begin{document}
\vspace*{1cm}
\begin{center}
{\LARGE \bf Inflationary cross-correlations of a non-minimal\\ \vskip 4pt  spectator and their soft limits}

\vspace*{1.0cm} 
{\large P. Jishnu Sai\footnote{\tt jishnup@iisc.ac.in}} and
{\large Rajeev Kumar Jain\footnote{\tt rkjain@iisc.ac.in}}\\
\vspace{.5cm} {\em Department of Physics, Indian Institute of Science\\
C. V. Raman Road, Bangalore 560012, India}\\
\end{center}
\vspace{8pt} 
\begin{abstract}
Light spectator fields may not be dynamically relevant for the inflationary phase of the early universe, but they can still induce interesting imprints on cosmological observables. In this paper, we compute the cross-correlations of the inflationary perturbations, both scalar and tensor, with the fluctuations of a non-minimally interacting spectator field using the in-in formalism and investigate the consistency relations associated with such cross-correlations. In particular, the scalar consistency relation is derived semi-classically by generalizing the consistency relation obtained earlier for cosmic magnetic fields.  Notably, we find that the direct coupling between the inflaton and the spectator solely determines the local non-linearity parameter associated with the scalar cross-correlation during slow-roll inflation, regardless of the specific form of the Lagrangian for the spectator field. Further, we calculate the tensor correlation with spectator fluctuations, explore the associated soft limits, and demonstrate the violation of the conventional tensor consistency relation with a non-minimal derivative coupling. Our analysis stresses that the violation of tensor consistency relations does not necessarily imply the superhorizon evolution of tensor modes. Instead, such violations can arise due to the non-minimal derivative coupling of the spectator field to gravity. 
Finally, we discuss the wider implications of our results in the context of cosmological soft theorems.
\end{abstract}


\section{Introduction}

The inflationary epoch in the very early universe provides a natural framework for understanding the large scale homogeneity and isotropy of our observed universe and the origin of primordial density perturbations which induce the temperature anisotropies in the cosmic microwave background (CMB) and later give rise to the formation of large scale structures in the universe \cite{Starobinsky:1980te, Guth:1980zm, Linde:1981mu, Mukhanov:1981xt, Bassett:2005xm, Sriramkumar:2009kg, Martin:2013tda}. Inflationary cosmology also presents itself as an interesting avenue to probe the primaeval interactions of quantum fields that may be dynamical during inflation but need not necessarily drive inflation. The quantum fluctuations of such fields are often assumed to be Gaussian in nature and, thus, completely described by the two-point correlation function or the power spectrum \cite{Martin:2013nzq, Planck:2018jri, Chowdhury:2019otk}. However, interactions among different fields or with gravity are only imprinted in the higher-order correlation functions and therefore, primordial non-Gaussianities (NG) are usually considered a novel measure of quantum interactions during inflation \cite{Gangui:1993tt, Gangui:1994yr, Gangui:1999vg, Wang:1999vf,  Bartolo:2004if, Lyth:2005fi, Seery:2005wm, Chen:2006nt, Chen:2010xka, Wang:2013zva, Kristiano:2021urj}. If the energy scale of inflation is very high, precise measurements of NG can provide interesting insights into quantum interactions at energy scales which are far beyond the reach of any laboratory experiments in the near future \cite{Planck:2019kim, LiteBIRD:2022cnt, Achucarro:2022qrl}. 

In order to probe the underlying physics of inflation and to gain further insights into the nature of primordial field interactions, the study of higher-order NG correlators has received enormous attention over the years within the community, leading to the development of numerous theorems and identities associated with these correlators \cite{Maldacena:2002vr, Creminelli:2004yq, Sloth:2006nu, Cheung:2007sv, Renaux-Petel:2010paw, Ganc:2010ff, Creminelli:2011rh, Chialva:2011hc, Schalm:2012pi, Pajer:2013ana, Senatore:2012wy, Creminelli:2012ed, Creminelli:2013cga, Berezhiani:2013ewa, Berezhiani:2014tda, Collins:2014fwa, Finelli:2017fml}. Among these, cosmological soft theorems hold particular significance which usually indicate a relation between an $(n + 1)$- and an $n$-point correlation function in the limit in which one of the modes is {\it soft}, i.e. its momentum is very small as compared to the others \cite{Giddings:2010nc, Giddings:2011zd, Hui:2018cag, Pajer-notes:2019}. These soft theorems are often related to a non-linearly realised symmetry of the action of cosmological perturbations, which might be spontaneously broken by the state of the underlying theory.
These symmetries are closely connected to cosmological adiabatic modes, which are crucial for understanding the statistics of primordial fluctuation in the early universe \cite{Weinberg:2003sw, Weinberg:2005vy, Hinterbichler:2012nm, Hinterbichler:2013dpa, Pajer:2017hmb}. 
An adiabatic mode is a cosmological perturbation that, on superhorizon scales, appears locally identical to a gauge mode and can be absorbed through a coordinate transformation \cite{Pajer-notes:2019, Mirbabayi:2014zpa}. 
Their existence is significant for various reasons, including constraining the number of degrees of freedom present during inflation and  their utility in deriving cosmological soft theorems.

One of the most well-known cosmological soft theorems in single field inflation is the Maldacena consistency relation (CR) \cite{Maldacena:2002vr}, which relates the bispectrum of the comoving curvature perturbation $\zeta$ to the power spectrum in the squeezed limit as
\beq
\lim_{k_1 \to 0}\frac{1}{P_{\zeta}(k_1)}\langle \zeta_{\vect k_1}\zeta_{\vect k_2}\zeta_{\vect k_3}\rangle  = -(2\pi)^3 \delta^{(3)}(\vect k_1+\vect k_2+\vect k_3)\;\left(\frac{\partial\ln [k_2^3P_\zeta]}{\partial \ln k_2}\right)P_{\zeta}(k_2),
\eeq
where $P_{\zeta}$ is the power spectrum.
This CR is derived using the background wave approach in which the long-wavelength $\zeta$ which is conserved outside the horizon, can be absorbed by an appropriate redefinition of coordinates. Similarly, one can also calculate the soft theorems associated with the bispectrum $\langle \gamma_{\vect k_1}\gamma_{\vect k_2}\gamma_{\vect k_3}\rangle$ of tensor perturbations $\gamma$, as well as the cross-correlations between the curvature and tensor perturbations such as $\langle \gamma_{\vect k_1}\zeta_{\vect k_2}\zeta_{\vect k_3}\rangle$ and $\langle \zeta_{\vect k_1}\gamma_{\vect k_2}\gamma_{\vect k_3}\rangle$ \cite{Maldacena:2011nz, Sreenath:2013xra, Sreenath:2014nca, Sreenath:2014nka}. 
In addition to the semi-classical Maldacena formalism,  cosmological soft theorems can be derived in different ways such as using Ward-Takahashi identities \cite{Assassi:2012zq, Hinterbichler:2012nm}, Slavnov-Taylor identities \cite{Berezhiani:2013ewa, Berezhiani:2014tda}, operator-product expansion \cite{Kehagias:2012pd}, and the wave functional technique \cite{Pimentel:2013gza}.

In effective field theories, it is natural to expect the presence of additional degrees of freedom during inflation besides the inflaton \cite{Adshead:2008gk, Achucarro:2022qrl}. 
If their impact on the overall background dynamics remains insignificant, they are commonly referred to as the \textit{light degrees of freedom}\footnote{The light degrees of freedom can be described by scalar fields, gauge fields or even fermions. They can be minimally or non-minimally coupled to gravity and can have various types of interactions with other light fields and the inflaton. Light scalar spectator fields are usually characterised by the small values of the ratio $m/H$.} or the spectator fields \cite{Kobayashi:2010fm, Vennin:2015egh, DeAngelis:2023fdu}. 
While the background expansion is driven by the energy density of the inflaton, the quantum fluctuations of the spectator during inflation leave remarkable imprints on the spectra of primordial perturbations. For this reason, imprints of light spectator fields on cosmological observables both during and after inflation
have been discussed extensively in the literature which can provide new insights into the physics of the early universe.
Some of the prominent examples of spectator fields are curvaton, axions, and primordial gauge fields. For instance, curvaton is a light scalar field during inflation which generates curvature perturbations at late times after the inflaton field has decayed. Moreover, the initial isocurvature perturbations of the curvaton are converted to the adiabatic 
curvature perturbation after inflation when the curvaton density becomes a significant fraction of the total energy density \cite{Enqvist:2001zp, Lyth:2001nq, Moroi:2001ct}. Besides being consistent with the power spectrum constraints from the CMB, the curvaton also induces a large amount of NG which makes it distinguishable from conventional single-field inflation \cite{Bartolo:2003jx, Sasaki:2006kq, Huang:2008ze, Kawasaki:2011pd, Enomoto:2012uy, Kawasaki:2012gg}. A broadly similar conclusion applies to most of the spectator scenarios that their NG signal is very different than single field models, however,  it crucially depends on the underlying dynamics of these models.

To study the bispectra of primordial perturbations in these models, it is important to study the cross-correlations between the fluctuations of such light fields and the inflationary scalar and tensor perturbations. Similar cross-correlations have already been explored in the literature, for the case of primordial gauge field with the inflationary curvature and tensor perturbations \cite{Caldwell:2011ra, Motta:2012rn, Jain:2012ga, Jain:2012vm, Shiraishi:2010kd, Shiraishi:2012xt, Chowdhury:2018blx, Chowdhury:2018mhj, Jain:2021pve}. Importantly, the squeezed limit of such correlators gives rise to a new set of CRs. Interestingly, ref. \cite{Jain:2012ga} proposed a new simple semi-classical derivation of the CR for the cross-correlation between the scalar metric perturbation and two powers of the magnetic field, for the kinetic coupling scenario that included a direct coupling between the inflaton and gauge field. One of the key inputs for this derivation was the inherent nature of the spectator field, specifically the conformal invariance of the gauge field in the absence of direct coupling which played a crucial role in deriving such CR\footnote{We are particularly indebted to Martin S. Sloth for sevaral valuable discussions and private communications on these topics.}.

To explore the significance of the nature of the spectator field in deriving such CRs, we consider a directly coupled light scalar spectator field $\sigma$, with the following Lagrangian 
\bea 
S_\sigma=\int d^4x \sqrt{-g}\,\lambda(\phi) \mathcal{L}_\sigma~,
\eea 
where $\lambda (\phi)$ is the direct coupling between the inflaton $\phi$ and $\sigma$. In the lower-dimensional effective UV complete theories, it is both expected and natural to encounter such direct couplings to the dilaton field or the moduli of the internal dimensions. Therefore, this direct coupling is also referred to as dilatonic coupling \cite{Taylor:1988nw}. Such direct couplings have also been studied in the models of inflationary magnetogenesis wherein $\mathcal{L}_\sigma$ is identified with the gauge field Lagrangian \cite{Turner:1987bw,  Ratra:1991bn, Durrer:2010mq, Byrnes:2011aa, Ferreira:2013sqa, Ferreira:2014hma, Ferreira:2014zia, Tripathy:2021sfb, Tripathy:2022iev}.
In general, we can choose $\mathcal{L}_\sigma$ in many different ways such as with minimal, non-minimal, conformal, non-conformal or even with derivative couplings. This motivates us to set up a model of non-minimally interacting spectator which also contains a non-trivial derivative coupling with gravity. We find that this derivative coupling plays a very crucial role in understanding the scalar and tensor CRs and the soft limits of their corresponding bispectra. 

In this paper, we study the inflationary correlation functions of the primordial curvature and tensor perturbations with the fluctuations of a non-minimally interacting scalar spectator field. Using the in-in formalism, we compute the full bispectrum of the scalar correlator and discuss in detail its squeezed limit. For the scalar cross correlation, the associated bispectrum in the squeezed limit is completely determined by the overall dilatonic coupling of the spectator field and does not depend on the explicit structure of its Lagrangian. Further, we derive a semi-classical CR associated with this scalar correlator and find that it agrees with the soft limit of the full bispectrum. We discuss various conditions under which the scalar CR can be violated. 
Besides the scalar bispectrum, we also compute the cross-correlation of the tensor mode with the spectator field and investigate its soft limit. Interestingly, the conventional tensor CR does not agree with the soft limit of the tensor cross-correlation. Usually, the violation of these CRs is associated with the non-adiabatic nature of the scalar and tensor fluctuations on superhorizon scales. However, in our case, we find that the violation arises due to the non-minimal derivative coupling of the spectator field. The violation of these fundamental CRs may also indicate a violation of the equivalence principle (EP) on cosmological scales. Thus, we use our setup to highlight the boundaries of the universality of the tensor CRs. 

This paper is organised as follows: In the following section,  we discuss our scenario for the spectator field and obtain the solutions for its Fourier modes. In section \ref{scalar-corr},  we compute the cross-correlation of the primordial curvature perturbation with the spectator perturbations and derive the associated soft theorem. In section \ref{tensor-corr}, we present a similar calculation for the tensor correlation, its soft limit and the corresponding soft theorem.  
Finally, in section \ref{conclusions}, we conclude our results and discuss the implications of scalar and tensor CRs. 
In the two appendices \ref{EM_exchange} and \ref{integrals}, we present the calculations of the energy-momentum exchange relation for the spectator and the evaluation of various integrals which appear in the scalar bispectrum, respectively. 

Throughout this paper, we work in natural units with $\hbar = c =1$, and the Planck mass $M_{\rm Pl}^2 =1/8\pi G$ is set to unity. Our metric convention is $(-,+,+,+)$.


\section{Dynamics of a non-minimally interacting spectator field}

It is crucial to employ certain simplified models to examine the subtleties of the cross-correlations between inflationary perturbations and spectator fields. In this section, we shall introduce a toy model for a non-minimally interacting spectator field $\sigma$. This particular set-up incorporates the non-minimal and derivative coupling of the spectator to gravity, through the Ricci scalar and the Ricci tensor, and a direct coupling with the inflaton field $\phi$. The action for $\sigma$ for such a scenario can be written as \cite{Sushkov:2009hk}
\begin{equation}
\label{action}
    S_{\sigma}=-\frac{1}{2}\int d^4x \sqrt{-g}\;\lambda(\phi)\left[\Big( g^{\mu \nu}+\alpha R^{\mu \nu}\Big) \partial_\mu \sigma \partial_\nu \sigma+2V(\sigma)+\frac{\xi}{6}R \sigma^2\right],
\end{equation}
where $\phi$ is the inflaton field and $\lambda(\phi)$ characterise a direct coupling between the inflaton and spectator. In the above action, $\xi$ and $\alpha$ are constants that indicate the strength of non-minimal and derivative coupling of $\sigma$ to gravity,  respectively and $V(\sigma)$ is the potential of the spectator. Moreover,  we limit ourselves to work with the quadratic potential, i.e., $V(\sigma)=\frac{1}{2}m^2\sigma^2$. Here, $R$ and $R_{\mu \nu}$ are the Ricci scalar and Ricci tensor which are straightforward to calculate using their respective definitions from the given metric of spacetime. The homogeneous and isotropic background during inflation is described by the spatially flat Friedmann-Lemaître-Robertson-Walker (FLRW) metric which is  given as, 
\bea
ds^2 = -dt^2 + a^2(t)\, d{\bf x}^2 = a^2(\tau)\left(-d\tau^2+d{\bf x}^2\right)~,
\eea
where $\tau$ is the conformal time, defined by $d\tau = dt/a$ and $a(\tau)$ is the scale factor. In the FLRW background, the Ricci scalar $R$ and non-zero components of $R_{\mu \nu}$ are, 
\bea
\label{Backgroud_Ricci}
R=6\left(\frac{a''}{a^3}\right),\;\;\;R_{00}=-3\left[\frac{a''}{a}-\left(\frac{a'}{a}\right)^2\right] ,\;\; R_{ij}=\delta_{ij} \left[\frac{a''}{a}+\left(\frac{a'}{a}\right)^2\right] ~.
\eea
Here, an overprime denotes a derivative with respect to $\tau$. 
The scale factor $a(\tau)$ is determined by the typical background dynamics associated with slow-roll inflation.

As mentioned in the introduction, we assume that the scalar spectator field $\sigma$ is light\footnote{For our scenario with non-trivial couplings present, we assume that the light spectator field does not contribute to the background energy density but only induces quantum fluctuations, which will evolve during the inflationary expansion.} and does not significantly affect the background dynamics. Therefore, we assume that the classical background value of the spectator field, $\sigma(t)$, is approximately zero. However, during inflation, quantum fluctuations can occur around this background value, denoted as $\sigma({\bf x},\tau)$. Ideally, one would denote it as $\delta \sigma$, but for notational convenience, we denote the fluctuations as $\sigma$ since the background is zero.
These fluctuations will evolve in the time-dependent background of inflation. Therefore, it is natural to expect a cross-correlation between these fluctuations and the inflationary perturbations. To explore these cross-correlations in our specific model, we can use the standard quantization procedure. This involves the mode expansion of $\sigma({\bf x},\tau)$ in the following manner,
\bea 
\label{mode_expn}
\sigma({\bf x},\tau)=\int \frac{d^3{\bf k}}{(2\pi)^3}\left[a_{{\bf k}}\sigma_k(\tau)+a^{\dagger}_{-{\bf k}} \sigma^*_k(\tau)\right]e^{i{\bf k}\cdot {\bf x}}~, 
\eea 
where $a_{{\bf k}}$ and $a^{\dagger}_{{\bf k}}$ are the annihilation and creation operators, respectively, and $\sigma_k(\tau)$ represents the mode function associated with momentum ${\bf k}$ at time $\tau$. The mode function $\sigma_k(\tau)$ obeys the classical equation of motion, which can be obtained by varying the action (\ref{action}) with respect to $\sigma$. We define the two-point correlation function of $\sigma$ in Fourier space as
\bea
\label{sigma_Pwrspctrm}
\langle\sigma({\bf k},\tau)\sigma({\bf k'},\tau)\rangle=(2\pi)^3\delta^{(3)} (\vect{k}+\vect{k'})P_{\sigma}(k,\tau)
\eea
where the power spectrum of $\sigma$ is simply given by $P_{\sigma}(k,\tau)=|\sigma_k (\tau)|^2$. 
To enhance the clarity of our analysis, let's first set $\alpha=0$ in eq. (\ref{action}) and vary the action, which leads to the equation of motion in the Fourier space as, 
\bea 
\label{eom1}
\sigma_k''+\left(2\frac{a'}{a}+\frac{\lambda'}{\lambda}\right)\sigma_k'+\left(k^2+a^2m^2+\xi \frac{a''}{a}\right)\sigma_k=0~.
\eea 
One can canonically normalize the field $\sigma$ by defining a variable $\tilde{\sigma}_k=aS\sigma_k$ with $S=\sqrt{\lambda}$ and recast the equation of motion in terms of $\tilde{\sigma}$ as a harmonic oscillator with a time-dependent mass term, 
\bea 
\label{Normalized_eom}
\tilde{\sigma}_k''+\left(k^2+a^2 m^2-(1-\xi)\frac{a''}{a}-\frac{S''}{S}-2\frac{a'S'}{a S}\right)\tilde{\sigma}_k=0~.
\eea 
It can be noted from the above equation that by setting $\xi=1$, $m=0$ and for a constant $\lambda$, the equation for $\tilde{\sigma}$ becomes identical to that of a massless scalar field in the Minkowski spacetime. This particular choice of parameters, known as the conformal coupling, causes $\tilde{\sigma}$ to exhibit a behaviour resembling that of a massless scalar field propagating in the flat spacetime. The direct coupling $\lambda(\phi)$ behaves as a dilatonic coupling which typically appears in effective field theories or scalar-tensor theories. To proceed further, we have to specify the functional form of $\lambda(\phi)$. We choose to work in the comoving gauge ($\delta \phi=0$), which allows us to express $\lambda$ as only a  function of time. Such a dilatonic coupling has also been studied in the scenarios of inflationary magnetogenesis as it breaks the conformal invariance of gauge fields which is a necessary condition to excite them during inflation. With these motivations, we can parameterise this direct coupling as a power law in the scale factor\footnote{In the context of inflationary magnetogenesis, a coupling function of the form $\lambda \propto a^{2n}$ with $n>0$ gives rise to the so-called strong coupling problem and there have been several works to address and resolve this problem \cite{Ferreira:2013sqa, Ferreira:2014hma, Tasinato:2014fia, BazrafshanMoghaddam:2017zgx}. In this work, we do not restrict ourselves to any particular choice of $n$, and thus, the strong coupling issue may simply be circumvented by avoiding the regimes with $n>0$, as was done, for instance, for gauge fields in Ref. \cite{Ferreira:2013sqa}. Moreover, this problem would not arise in the first place if $\sigma$ is a scalar field associated with the dark sector which is not necessarily coupled with the standard model.}, i.e., $\lambda \propto a^{2n}$. Then the effective mass term in eq. (\ref{Normalized_eom}) becomes, 
\bea 
\label{m_eff}
a^2 m^2-(1-\xi)\frac{a''}{a}-\frac{S''}{S}-2\frac{a'S'}{a S}=-\frac{1}{\tau^2}\left[2(1-\xi)+n (n+1)+2n-\left(\frac{m^2}{H^2}\right)\right]~, 
\eea
where we have used $a(\tau) \simeq -1/(H\tau)$ during inflation. With this result, we find that the solution of the differential equation (\ref{Normalized_eom}) can be written in terms of Hankel functions. Assuming the standard Bunch-Davis initial conditions for the Fourier modes, we obtain the following solution for $\tilde{\sigma}$ as,
\bea
\tilde{\sigma}(k,\tau)=\frac{\sqrt{\pi}}{2}\;e^{i(\nu+1/2)\pi/2}\sqrt{-\tau}H^{(1)}_\nu(-k\tau)~,
\eea
where $H^{(1)}_\nu (x)$ denotes the Hankel function of the first kind of order $\nu= \sqrt{(n+3/2)^2-2\xi-(m^2/H^2)}$.
With this, the mode function solution is 
\bea
\label{modefn0}
\sigma (k,\tau)=\frac{\sqrt{\pi}}{2}\;e^{i(\nu+1/2)\pi/2}\frac{\sqrt{-\tau}}{a\sqrt{\lambda}}H^{(1)}_\nu(-k\tau)~.
\eea
The time dependence in the mode function comes from the terms outside the Hankel function, which scale as $\sqrt{-\tau}/a\sqrt{\lambda}\sim \tau^{n+3/2}$, whereas the Hankel function in the superhorizon limit $(|k\tau| \ll 1)$ scale as $\sim\tau^{-\nu}$.
It is important to note that other terms do not cancel the superhorizon scaling of the Hankel function in general. Thus the Fourier modes will evolve on superhorizon scales and will only freeze for a particular choice of parameters, i.e., $\xi=0,m=0$ or $\xi=-m^2/2H^2$.

Now, let us consider the action (\ref{action}) with a non-zero  $\alpha$ and vary it to derive the equation of motion which leads to,
\begin{equation}
\label{Modfn}
\sigma_k''+\left(2\frac{a'}{a}+\frac{\lambda'}{\lambda}\right)\sigma_k'+\left(k^2+\frac{a^2m^2}{1+3\alpha H^2}+\frac{\xi}{1+3\alpha H^2}\frac{a''}{a}\right)\sigma_k=0~.
\end{equation}
To arrive at the above expression, we have used the scale factor $a \simeq -1/(H \tau)$ in eq. (\ref{Backgroud_Ricci}) and obtained components of the background Ricci tensor as $ R_{00}=-3a^2H^2$ and $R_{ij}=3a^2H^2 \delta_{ij}
$. By comparing this equation with eq. (\ref{eom1}), we observe that they are identical under the redefinition of the following  parameters as,
\bea
m^2 \to \tilde{m}^2 = \frac{m^2}{1+3\alpha H^2},\quad \text{and}\quad \xi \to \tilde{\xi} =\frac{\xi}{1+3\alpha H^2}\;.
\eea 
This indicates that we can obtain the solution of eq. (\ref{Modfn}) in the same manner as earlier. But in this case, the canonically normalized field will be $\tilde{\sigma}_k=\sigma_ka\sqrt{(1+3\alpha H^2)\lambda} $ and the complete solution for the mode function would be,
\bea
\label{Rij_modefn}
\sigma (k,\tau)=\frac{1}{\sqrt{1+3\alpha H^2}}\frac{\sqrt{\pi}}{2}\;e^{i(\nu+1/2)\pi/2}\frac{\sqrt{-\tau}}{a\sqrt{\lambda}}H^{(1)}_\nu(-k\tau)~,
\eea
with  $\nu=\sqrt{(n+3/2)^2-2\tilde{\xi}-(\tilde{m}^2/H^2)}$. Also, in this case, the modes will evolve on superhorizon scales but will remain frozen for $\tilde{\xi}=0,\tilde{m}=0$ or $ \tilde{\xi}=-{\tilde{m}}^2/2H^2$, as earlier, even for the case when the derivative coupling is present and non-vanishing, i.e., $\alpha \neq 0$. 

\section{Cross-correlation with curvature perturbation and the consistency relation}
\label{scalar-corr}
In this section, we shall discuss our calculations of the three-point cross-correlation of the comoving curvature perturbation with the fluctuations of the spectator field. To do so, we use the in-in formalism, which is a standard framework for studying equal-time quantum correlations in the early universe. In this formalism, the interaction Hamiltonian plays a crucial role in capturing the effects of interactions between different fields. For our case, we focus on the interaction Hamiltonian $H_{\zeta\sigma\sigma}$, which describes the coupling between the curvature perturbation $\zeta$ and the spectator field $\sigma$. Thus, the cubic order interaction Hamiltonian can be constructed as follows, 
\bea 
\label{Gen_H}
H_{\zeta\sigma\sigma}=-\frac{1}{2}\int d^3x \;\sqrt{-g} \;T^{\mu \nu}\delta g_{\mu \nu}~,
\eea
where $T^{\mu \nu}$ represents the stress-energy tensor of the spectator $\sigma$. For a systematic analysis of the dynamics of metric perturbations at the action level, we use the standard Arnowitt-Deser-Misner (ADM) parametrisation of the metric as,
 \bea 
 ds^2=-N^2dt^2+h_{ij}(dx^i+N^i dt)(dx^j+N^jdt)~, 
 \eea 
  where $N({\bf x},t)$ and $N^i({\bf x},t)$ are called the lapse function and the shift vector, respectively. The dynamical degrees of freedom are contained in the spatial part of the metric $h_{ij }$  whereas lapse and shift are like the Lagrangian multipliers which are determined by the constraint equations. In the work, we mostly work in the comoving gauge where $
 \delta\phi=0$ and and spatial metric is parameterised as $h_{ij}=a^2 e^{2\zeta}[e^\gamma]_{ij}$. In this gauge, the first-order constraint equations give, 
\bea 
N=1+\frac{\dot{\zeta}}{H},\;\; N_i=\p_i\left(-\frac{\zeta}{H}+\ep a^2\p^{-2}\dot \zeta \right)~, 
\eea 
 where the overdot denotes the time derivative with respect to $t$, $\epsilon$ is the first slow-roll parameter, and $\partial^{-2}$ denotes the inverse Laplacian operator. Then the metric perturbations at the first order are, 
\bea
\label{dGs}
\delta g_{00}=-2\frac{\dot{\zeta}}{H}, \; 
\delta g_{0i}=\p_i\left(-\frac{\zeta}{H}+\ep a^2\p^{-2}\dot \zeta \right), \; \text{and}\;\;
\delta g_{ij}=2a^2\zeta \delta_{ij}~.
\eea
 Using eq. (\ref{dGs}) in eq. (\ref{Gen_H}), and by performing some trivial integration by parts, we obtain the following interaction Hamiltonian 
\bea
\label{NM_EM}
H_{\zeta \sigma \sigma}=-\int d^3x\; a^3 \frac{\zeta}{H}  \left(\nabla_\mu T^{\mu 0}\right)+\mathcal{O}(\ep)~.
\eea
 This indicates that if the four divergences of the energy-momentum tensor of the spectator field are zero, the cubic order interaction Hamiltonian is slow-roll suppressed. But, in our case, the spectator is directly coupled to the inflaton. As a result, there will be energy momentum exchange between the inflaton fluctuations and the spectator field. So the divergence of the energy-momentum tensor of the spectator will not be zero, and it can be calculated,
\bea
\label{energy_exchange}
\nabla_\mu T^{\mu \nu}=-\frac{1}{2}\nabla^\nu \lambda \left( \left(g^{\rho \kappa}+\alpha R^{\rho\kappa}\right) \partial_\rho \sigma \partial_\kappa \sigma+2V(\sigma)+\frac{\xi}{6}R \sigma^2\right)~. 
\eea
This expression can be trivially obtained by invoking the diffeomorphic invariance of the action (\ref{action}), and it is outlined in detail in appendix {\ref{EM_exchange}}. To proceed further,  we first set $\alpha=0$ to ensure clarity and introduce it later, as we did in the previous section. Using the above equation in eq. (\ref{NM_EM}) and rewrite the leading order interaction Hamiltonian in terms of conformal time as follows, 
\bea 
H_{\zeta \sigma \sigma}=-\frac{1}{2}\int d^3x a^2\lambda'(\tau)\tau\zeta\left(\sigma'^2-(\p\sigma)^2-\left(a^2m^2+\xi\frac{a''}{a}\right)\sigma^2\right)~.
\eea
At this level, we have dropped all the terms that are proportional to slow roll parameters. As in (\ref{mode_expn}), curvature  perturbation $\zeta$ is also mode expanded and the corresponding mode function obtained in the standard manner with the solution, 
\bea
\zeta_k(\tau)=\frac{1}{\sqrt{2\epsilon}}\frac{H}{\sqrt{2k^3}} \left(1+ik\tau\right)e^{-ik\tau}~.
\eea 
Using this interaction Hamiltonian in the in-in master formula for a three-point function $\mathcal{O}$, 
\bea 
\label{in_in}
\lle\mathcal{O}(\tau)\rgr= -i\int ^\tau d\tau' \lle\big[\mathcal{O}(\tau), H_{\zeta \sigma \sigma}(\tau' )\big] \rgr
\eea 
we calculate the three-point correlator $\lle \zeta \sigma \sigma\rgr$ and obtain, 
\bea 
\label{FIN}
\lle \zeta({\bf k}_1,\tau_I) \sigma ({\bf k}_2,\tau_I) \sigma ({\bf k}_3,\tau_I) \rgr=(2\pi)^3\delta^{(3)}({\bf k}_1+{\bf k}_2+{\bf k}_3)\left[-\left(\mathcal{I}_1+{\bf k}_2\cdot{\bf k}_3\,\mathcal{I}_2\right)+\left(\frac{m^2}{H^2}+2\xi\right)\mathcal{I}_3\right],
\eea 
with, 
\bea
\mathcal{I}_1&=&2\,{\rm Im} \left[\zeta_{k_1}(\tau_I)\sigma_{k_2}(\tau_I)\sigma_{k_3}(\tau_I)\int d\tau \tau a^2 \lambda'(\tau) \zeta_{k_1}^*(\tau){\sigma'}_{k_2}^*(\tau){\sigma'}_{k_3}^*(\tau)\right]\label{I1}~, \\
\mathcal{I}_2 &=&2\,{\rm Im} \left[\zeta_{k_1}(\tau_I)\sigma_{k_2}(\tau_I)\sigma_{k_3}(\tau_I)\int d\tau \tau a^2 \lambda'(\tau) \zeta_{k_1}^*(\tau)\sigma_{k_2}^*(\tau)\sigma_{k_3}^*(\tau)\right]\label{I2}~,\\
\mathcal{I}_3&=&2\,{\rm Im} \left[\zeta_{k_1}(\tau_I)\sigma_{k_2}(\tau_I)\sigma_{k_3}(\tau_I)\int \frac{d\tau}{\tau} a^2 \lambda'(\tau) \zeta_{k_1}^*(\tau){\sigma}_{k_2}^*(\tau){\sigma}_{k_3}^*(\tau)\right]\label{I3}~. 
\eea
The result in eq. (\ref{FIN}) is our complete result for the correlator and the integrals can be evaluated for the most general case. 
Moreover, to study the CR associated with $\lle \zeta \sigma \sigma\rgr$, we have to evaluate the integrals (\ref{I1}), (\ref{I2}) and (\ref{I3}) using the explicit form of the coupling function, i.e., $\lambda (\tau) \propto a^{2n} \propto \tau^{-2n}$ and the mode functions. However, in the squeezed limit, i.e. ${\bf k}_1 \to 0$, and ${\bf k}_2\simeq -{\bf k}_3$, we can show that
\bea 
\label{I123SL}
\mathcal{I}_1=-(2n)\; |\zeta_{k_1}(\tau_I)|^2 |\sigma_{k_2}(\tau_I)|^2+k_2^2\mathcal{I}_2+\left(2\xi+\frac{m^2}{H^2}\right)\mathcal{I}_3~.
\eea
The detailed derivation of this equation is given in appendix \ref{integrals}. We have arrived at it by using the equation of motion of mode function and the normalized Wronskian.
By using eq. (\ref{I123SL}) in the squeezed limit of eq. (\ref{FIN}), we get the squeezed limit correlator as
\bea 
\label{sqzd_L1}
\lim_{{\bf k}_1 \to 0}
\lle \zeta({\bf k}_1,\tau_I) \sigma ({\bf k}_2,\tau_I) \sigma ({\bf k}_3,\tau_I) \rgr
=2n\;(2\pi)^3\delta^{(3)}({\bf k}_1+{\bf k}_2+{\bf k}_3) P_{\zeta}(k_1)P_{\sigma}(k_2)~. 
\eea 
This relation is precisely the form of the CR for this correlator, i.e., the three-point cross-correlation in the squeezed limit is proportional to the product of two power spectra. The strength of the local non-linearity parameter associated with this correlator is simply  $2n$ which can be expressed in terms of the direct coupling as follows,  
\bea 
\frac{d\ln \lambda}{d\ln a}=\frac{\dot{\lambda}}{H \lambda}=2n~.
\eea 
It is remarkable that the non-linearity parameter is independent of the other parameters of our model, i.e., for any $\xi$ and $m$, we get the same CR as in eq. (\ref{sqzd_L1}). Therefore, our CR holds true regardless of whether the spectator field is conformal or non-conformal, massive or massless. This observation highlights the general applicability of the new CR, suggesting its wider scope and validity across different scenarios and field properties.

To verify the generality of this CR within our toy model, we can now consider the case with non-minimal derivative coupling ($\alpha \neq 0$). It is trivial to compute the interaction Hamiltonian with the non-minimal derivative coupling using (\ref{NM_EM}) and (\ref{energy_exchange}). Using this interaction Hamiltonian in the in-in master formula (\ref{in_in}), we obtain the correlator as, 
\bea 
\label{FIN_D}
\lle \zeta({\bf k}_1,\tau_I) \sigma ({\bf k}_2,\tau_I) \sigma ({\bf k}_3,\tau_I) \rgr=(2\pi)^3\delta^{(3)}({\bf k}_1+{\bf k}_2+{\bf k}_3)\left[-(1+3\alpha H^2)\left(\mathcal{I}_1+{\bf k}_2\cdot{\bf k}_3\mathcal{I}_2\right)+\left(\frac{m^2}{H^2}+2\xi\right)\mathcal{I}_3\right].
\eea
Here the integrals $\mathcal{I}_1$, $\mathcal{I}_2$, and $\mathcal{I}_3$ refer to the same integrals defined in equations (\ref{I1}), (\ref{I2}), and (\ref{I3}). However, it is important to note that in this case, the mode function given in eq. (\ref{Rij_modefn}) should be used. Now let's compute the squeezed limit of the above correlator for which we use the squeezed limit of $\mathcal{I}_1$, $\mathcal{I}_2$, and $\mathcal{I}_3$. As discussed in appendix \ref{integrals}, even for $\alpha \neq 0$ in the squeezed limit, the integral $\mathcal{I}_1$ can be expressed in terms of $\mathcal{I}_2$ and $\mathcal{I}_3$, similar to eq. (\ref{integral-relation}). 
Interestingly, even for this scenario with $\alpha \neq 0$, we obtain precisely the same CR as in eq. (\ref{sqzd_L1}). Consequently, we have also derived the squeezed limit of the cross-correlation of curvature perturbation with the non-minimally interacting spectator field (\ref{action}) at the leading order in slow-roll parameters. The resulting expression is given by, 
\bea 
\label{zss_squeezed}
\lim_{{\bf k}_1 \to 0}
\lle \zeta({\bf k}_1,\tau_I) \sigma ({\bf k}_2,\tau_I) \sigma ({\bf k}_3,\tau_I) \rgr
=\frac{\dot{\lambda}}{H\lambda}\;(2\pi)^3\delta^{(3)}({\bf k}_1+{\bf k}_2+{\bf k}_3) P_{\zeta}(k_1)P_{\sigma}(k_2)~. 
\eea 
This result bears a striking resemblance with the squeezed limit correlator of the curvature perturbation with gauge fields as found in \cite{Jain:2012ga, Jain:2012vm}, which is not a mere coincidence\footnote{One may notice a sign difference in the non-linearity parameter in eq. (\ref{zss_squeezed}) and ref. \cite{Jain:2012vm}. We recently found that there is actually a typo in \cite{Jain:2012ga} which is also there in the results of \cite{Jain:2012vm}.}. In both cases, a light degree of freedom is directly coupled to the inflaton, and this direct coupling is realized as a power law in the scale factor. 

From the above consistency relation, we observe that the corresponding local non-linearity parameter is independent of the parameters $m$, $\xi$ and $\alpha$ as in the earlier case. This is an interesting and a non-trivial result. The factor $\dot{\lambda}/(H \lambda) $ can be understood as the new scale introduced by the direct coupling $\lambda$ measured in the units of Hubble parameter. Since we are naturally working in the regime $\dot{\lambda}/(H \lambda) \gg \epsilon, \eta$ where $\epsilon$ and $\eta$ are the slow roll parameters, one can heuristically estimate the strength of the leading order interaction Hamiltonian as $H_{\zeta \sigma \sigma}/H_{\sigma \sigma} \propto \dot{\lambda}/(H \lambda) \cdot P_\zeta^{1/2}$, similar to \cite{Jain:2012ga}. Consequently, we expect the local non-linearity parameter to be of the order of  $\dot{\lambda}/(H \lambda)$. Moreover, a careful look at the cubic order action (\ref{cubic_action}) provides us the insight that when $\zeta$ becomes superhorizon and frozen, the action is identical to the quadratic action of $\sigma$ but with a new direct coupling $-\dot{\lambda}\zeta/H$. It means that we can determine  $\lle \zeta_L {\lle \sigma \sigma \rgr}_{\zeta_L}\rgr$  if we know the two point correlator of $\sigma$ for an arbitrary coupling function $\lambda$. This observation allow us to derive the above CR\footnote{In the literature, CRs and soft theorems are often used interchangeably. It is important to note that while CRs represent model-independent statements about cosmological observables, soft theorems are a subset of CRs arising from the non-linear realisation of symmetries within cosmological correlators.} semi-classically and understand as to why the local non-linearity parameter is purely determined by the coupling function. To do so, let's work in the flat gauge. One can easily see that the lapse and shifts are proportional to slow roll parameters, as shown in eq. (2.24) of \cite{Maldacena:2002vr}. Then, the third order action can be written as, 
\bea
S^{(3)}=
 -\frac{1}{2}\int d^4x \sqrt{-g}\;\partial_\phi\lambda \;
 \delta\phi\left[ \left(g^{\mu \nu} +\alpha R^{\mu \nu}\right)\partial_\mu \sigma \partial_\nu \sigma+2V(\sigma)+\frac{\xi}{6}R \sigma^2\right]+ \text{slow-roll suppressed terms}. 
\eea
This action is consistent with the result we obtained when we appropriately translated eq. (\ref{NM_EM}) from the comoving gauge to the flat gauge. It is evident that the influence of inflaton fluctuation solely comes through the coupling $\lambda (\phi)$ when we drop the slow-roll suppressed terms. Consequently, it becomes apparent that the long-wavelength inflaton fluctuations can be absorbed into the coupling function, resulting in the action being transformed back to a second-order form but with a modified coupling. This allows us to study the effects of the long wavelength perturbation $\delta\phi_L$ on the short wavelength fluctuations of $\sigma$ by defining the effective coupling as $
\lambda_B=\lambda(\phi_0+\delta\phi_L)=\lambda_0+\partial_\phi\lambda \delta \phi_L
$. Using this expansion, the two-point correlator of the spectator field with the modified coupling can be written as, 
\bea
\lle \sigma \sigma\rgr_B=\lle \sigma \sigma\rgr_0+\frac{\partial\lle \sigma \sigma\rgr_B}{\partial\delta\phi_L}\bigg\rvert_{\delta\phi_L=0}\delta\phi_L+\cdots
\eea 
It is now straightforward to find the squeezed correlator of the form $\lle \delta\phi_L\lle \sigma \sigma\rgr_B\rgr$ as analogous to the approach of Maldacena CR. For this, one has to evaluate $\lle \sigma \sigma\rgr$ for a given form of $\lambda(\phi)$, and in our case, the coupling function takes the form $\lambda (\phi)=e^{2\phi/M}$, then $\lambda_B$ is just a rescaled $\lambda_0$ by a constant factor $1+\delta\phi_L\partial_\phi\lambda/\lambda$. By observing the fact that eq. (\ref{Normalized_eom}) is insensitive to a constant rescaling of $\lambda$, we can write the correlator in the coordinate space as
\bea 
\lle \sigma({\bf x}_1,\tau) \sigma({\bf x}_2,\tau) \rgr_B=\lle\frac{1}{a^2 \lambda_B}\tilde {\sigma}({\bf x}_1,\tau) \tilde {\sigma} ({\bf x}_2,\tau) \rgr \approx \lle \sigma({\bf x}_1,\tau) \sigma({\bf x}_2,\tau)\rgr_0-\delta\phi_L\frac{\partial_\phi\lambda} {\lambda} \lle \sigma({\bf x}_1,\tau) \sigma({\bf x}_2,\tau)\rgr_0.  
\eea
This two-point function can further be correlated with $\delta \phi$ which, in the Fourier space, will lead to, 
\bea 
\lim_{{\bf k}_1 \to 0}
\lle \delta\phi({\bf k}_1,\tau_I) \sigma ({\bf k}_2,\tau_I) \sigma ({\bf k}_3,\tau_I) \rgr
=-\frac{\partial_\phi\lambda}{\lambda}\;(2\pi)^3\delta^{(3)}({\bf k}_1+{\bf k}_2+{\bf k}_3) P_{\delta\phi}(k_1)P_{\sigma}(k_2)~.
\eea 
Note that, the above result is obtained in the flat gauge. We can translate it into the comoving gauge by using $\sqrt{2\epsilon}\,\zeta=\delta\phi$ and also using $\sqrt{2\epsilon}\,\partial_\phi\lambda=-\dot{\lambda}/H$, to find  
\bea 
\lim_{{\bf k}_1 \to 0}
\lle \zeta({\bf k}_1,\tau_I) \sigma ({\bf k}_2,\tau_I) \sigma ({\bf k}_3,\tau_I) \rgr
=\frac{\dot{\lambda}}{\lambda H}\;(2\pi)^3\delta^{(3)}({\bf k}_1+{\bf k}_2+{\bf k}_3) P_{\zeta}(k_1)P_{\sigma}(k_2)~, 
\eea  
where $\dot{\lambda}/H\lambda=2n$ for $\lambda \propto a^{2n}$. Here, we would like to stress that in ref. \cite{Jain:2012ga}, the CR is derived in the comoving gauge by using the conformal nature of the gauge field. In comparison, our result is independent of such conformal nature of the spectator field. This CR can be interpreted as a simple yet non-trivial consequence of direct coupling. 

Before we close this section, let us make a few subtle remarks about the conditions under which the background wave method (Maldacena approach) works. It is well known that the underlying working principle of the Maldacena CR is based on constructing an adiabatic mode inside the Hubble radius. In inflation, an adiabatic mode is a long wavelength frozen mode that is indistinguishable from a coordinate transformation. Using this approach, the CR for our scenario can be written as, 
\bea 
\lim_{{\bf k}_1 \to 0} \frac{1}{P_\zeta(k_1)}
\lle \zeta_{{\bf k}_1}\sigma_ {{\bf k}_2} \sigma _{{\bf k}_3}\rgr
=-(2\pi)^3\delta^{(3)}({\bf k}_1+{\bf k}_2+{\bf k}_3)\frac{\partial\ln[k_2^3P_\sigma(k_2)]}{\partial \ln k_2} P_\sigma(k_2)~.
\eea 
In the superhorizon limit, we find that $k^3P_\sigma \sim k^{3-2\nu}$ and $\nu$ is given after eq. (\ref{Rij_modefn}). Therefore, this CR is in agreement with (\ref{zss_squeezed}) {\it only} when $2\xi+m^2/H^2=0$. In other words, the CR obtained using the background wave method aligns with the soft limit of the in-in result exclusively when the spectator $\sigma$ becomes frozen in the superhorizon limit. However, it is noteworthy that the CR obtained using our semi-classical approach remains applicable within this setup without such limitations.
For scenarios involving direct dilatonic coupling, our analysis shows that all the non-minimal gravitational interactions are suppressed by the slow-roll parameters, and they can be ignored in the slow-roll limit. Furthermore, due to the evolution of spectator modes on superhorizon scales (with exceptions for specific parameter choices), one can not apply the Maldacena formalism to derive the CR. Therefore, one has to resort to our semi-classical formalism to obtain the CR for the cross-correlation of the curvature perturbation with spectator fields.

\section{Tensor cross-correlation and the consistency relation}
\label{tensor-corr}
In this section, we shall compute the tensor cross-correlation with our non-minimal spectator and study its squeezed limit. Contrary to the scalar CRs, it has been discussed that the tensor CRs are more robust conditions which are preserved in most situations and reflect the adiabatic nature of tensor modes during inflation. These CRs can only be violated in specific situations. It is observed that tensor CRs remain valid even when there are multiple scalar fields as long as any anisotropies decrease rapidly in an exponential manner \cite{Bordin:2016ruc}. This is because, in an expanding universe with  anisotropies decreasing rapidly in an exponential manner, the graviton mode 
 becomes constant on superhorizon scales. Therefore, there exists an adiabatic tensor mode that is locally indistinguishable from a pure gauge mode i.e. it can be absorbed by means of a suitable coordinate transformation and used to derive the conventional tensor CRs.  
 
 Here, we explicitly compute the cross-correlation of spectator fluctuations with the tensor perturbation using in-in formalism within our toy model and study the robustness of tensor CR. To perform our calculation, we have to mode expand tensor perturbations. Using the standard quantisation formalism, the mode expansion for tensor perturbations is defined as follows,
 \bea
\label{gamma_ij}
\gamma_{ij}({\bf x},\tau) 
= \int\frac{d^{3}{\bf k}}{(2\pi)^{3}}\sum_{s = \pm 2}
\left[\gamma_{k}(\tau) \, e^{i{\bf k}\cdot {\bf x}} \,  \epsilon ^{s}_{ij}(\hat{\bf k}) \, b_{\bf k}^s + h.c. \right],
\eea
where $\epsilon_{ij}^{s}$ represents the polarization tensor corresponding to helicity $s$. The normalization condition is given by $\epsilon_{ij}^{s} \epsilon_{ij}^{*s'}=2\delta_{ss'}$. The creation and annihilation operators satisfy the usual commutation relation $[b_{\bf k}^s,b^{{s'\dagger}}_{\bf k'}]=(2\pi)^3 \delta^{(3)}({\bf k}-{\bf k'}) \delta{s s'}$.
 The amplitude of the tensor mode during inflation is obtained in the standard manner as, 
\bea 
\gamma_k(\tau)=\frac{H}{\sqrt{k^3}} \left(1+ik\tau\right)e^{-ik\tau}~.
\eea 
Note that, we often suppress the helicity index $s$ since we do not have any parity violating term. To proceed further, it is necessary to obtain the Ricci tensor up to the first order in tensor perturbations ($\gamma_{ij}$), which is given by
\bea 
R_{ij}=\delta_{ij} \left[\frac{a''}{a}+\left(\frac{a'}{a}\right)^2\right]+\gamma_{ij}\left[\frac{a''}{a}-\left(\frac{a'}{a}\right)^2\right]~.
\eea 
During inflation, we can write $a''/a \simeq 2a^2H^2$ and $a'/a =aH$.
Then the cubic order Lagrangian for $\lle\gamma\sigma \sigma\rgr$ can be trivially derived from eq. (\ref{action}). Therefore, the corresponding interaction Hamiltonian can be written as, 
\bea 
H_{\gamma \sigma \sigma}=-\frac{1}{2}\int d^3xa^2\lambda\left(1-\alpha H^2\right)\gamma_{ij} \p_i\sigma \p_j\sigma~.
\eea 
Similar to the earlier section, in this case as well, we assume the power law parameterisation for the direct coupling, i.e., $\lambda \propto a^{2n} \propto \tau^{-2n}$. Using the above expression for the interaction Hamiltonian in the master in-in formula (\ref{in_in}), we obtain the tensor cross-correlation as
\bea 
\label{InIn_tensor}
\lle \gamma({\bf k}_1,\tau_I) \sigma ({\bf k}_2,\tau_I) \sigma ({\bf k}_3,\tau_I) \rgr&=&2\,(2\pi)^3\delta^{(3)}({\bf k}_1+{\bf k}_2+{\bf k}_3) \ep_{ij}k_{2i}k_{3j}(1-\alpha H^2)\nonumber\\ &\times&{\rm Im} \left[\gamma_{k_1}(\tau_I)\sigma_{k_2}(\tau_I)\sigma_{k_3}(\tau_I)\int d\tau a^2 \lambda(\tau) \gamma_{k_1}^*(\tau)\sigma_{k_2}^*(\tau)\sigma_{k_3}^*(\tau)\right] ~.
\eea 
The integral involved in the above equation is analogous to the integral $\mathcal{I}_2$, which can be evaluated similarly as in appendix \ref{integrals} but using the mode function $\sigma$ as in eq. (\ref{Rij_modefn}) for $\alpha \neq 0$. To investigate the tensor CR, let us consider the squeezed limit ${\bf k}_1 \to 0$, and ${\bf k}_2\simeq -{\bf k}_3$. Then, we can write the integral in (\ref{InIn_tensor}) similar to (\ref{app_I2}) as
\bea 
{\rm Im} \left[\gamma_{k_1}(\tau_I)\sigma_{k_2}(\tau_I)\sigma_{k_3}(\tau_I)\int d\tau a^2 \lambda(\tau) \gamma_{k_1}^*(\tau)\sigma_{k_2}^*(\tau)\sigma_{k_3}^*(\tau)\right]=-|\gamma_{k_1}(\tau_I)|^2|\sigma_{k_2}(\tau_I)|^2 \nonumber \\ \qquad  \times \; {\rm Im} \left[e^{i(\nu+1/2)\pi}\int d\tau  a^2 \lambda(\tau) \left(\sigma_{k_2}^*(\tau)\right)^2\right] ~.
\eea 
Upon using the mode function (\ref{Rij_modefn}), the right hand side of the above equation takes the following form,  
\bea 
\frac{\pi}{4(1+3\alpha H^2)} |\gamma_{k_1}(\tau_I)|^2|\sigma_{k_2}(\tau_I)|^2{\rm Im} \left[\int d\tau \tau  \left(H_{\nu}^{(2)}(-k_2\tau)\right)^2\right]~,
\eea
where the factor $1/(1+3\alpha H^2)$ evidently appears from the normalisation of the mode function, as given in  (\ref{Rij_modefn}).  This can be quickly evaluated using the eq. (\ref{Idenity_H2}) and we get, 
\bea 
{\rm Im} \left[\gamma_{k_1}(\tau_I)\sigma_{k_2}(\tau_I)\sigma_{k_3}(\tau_I)\int d\tau a^2 \lambda(\tau) \gamma_{k_1}^*(\tau)\sigma_{k_2}^*(\tau)\sigma_{k_3}^*(\tau)\right]=-\frac{1}{(1+3\alpha H^2)} \frac{\nu}{2k^2}|\gamma_{k_1}(\tau_I)|^2|\sigma_{k_2}(\tau_I)|^2
\eea
where $\nu$ is the order of the Hankel function appearing in (\ref{Rij_modefn}). With this, the final result for the correlator in the squeezed limit, i.e. ${\bf k}_1 \to 0$, and ${\bf k}_2\simeq -{\bf k}_3\equiv {\bf k}$, is
\bea 
\label{sqzd_tss}
\lim_{{\bf k}_1 \to 0}
\lle \gamma({\bf k}_1,\tau_I) \sigma ({\bf k}_2,\tau_I) \sigma ({\bf k}_3,\tau_I) \rgr=(2\pi)^3\delta^{(3)}({\bf k}_1+{\bf k}_2+{\bf k}_3)\ep_{ij}\frac{k_{i}k_{j}}{k^2}\left(\frac{1-\alpha H^2}{1+3\alpha H^2}\right)\nu P_\gamma(k_1) P_\sigma(k)~.
\eea 
For those acquainted with the conventional background wave method, it is evident that such methods can only capture the result for the case without the derivative coupling. To illustrate it further, let's write eq. (\ref{sqzd_tss}) for $\alpha=0$,
\bea 
\lim_{{\bf k}_1 \to 0}
\lle \gamma({\bf k}_1,\tau_I) \sigma ({\bf k}_2,\tau_I) \sigma ({\bf k}_3,\tau_I) \rgr=(2\pi)^3\delta^{(3)}({\bf k}_1+{\bf k}_2+{\bf k}_3)\ep_{ij}\frac{k_{i}k_{j}}{k^2}\nu P_\gamma(k_1) P_\sigma(k)~.
\eea 
On the other hand, the semi-classical derivation using the background wave approach gives, 
\bea 
\lim_{{\bf k}_1 \to 0} \frac{1}{P_\gamma(k_1)}
\lle \gamma_{{\bf k}_1}\sigma_ {{\bf k}_2} \sigma _{{\bf k}_3}\rgr
=-(2\pi)^3\delta^{(3)}({\bf k}_1+{\bf k}_2+{\bf k}_3)\ep_{ij}\frac{k_{i}k_{j}}{k^2}\frac{\partial}{\partial \ln k^2} P_\sigma(k)
\eea 
It can be derived from eq. (\ref{Rij_modefn}) that the power spectrum of the spectator field in the superhorizon limit scales as $P_\sigma \sim k^{-2\nu}$ and the derivative term in the above equation gives $\frac{\partial P_\sigma}{\partial \ln k^2} =-\nu P_\sigma $. 
This clearly indicates that both in-in and the semi-classical results are in agreement with each other {\it only} for $\alpha=0$, but they disagree for $\alpha \neq 0$. In the limit $\alpha=0$,  $\nu= \sqrt{(n+3/2)^2-2\xi-(m^2/H^2)}$ and thus, we observe that, for the two approaches to be in agreement, we only require $\alpha = 0$ irrespective of $\xi$ and $m$.
This shows the universality of tensor CRs. However, our analysis shows a violation of tensor CRs in the presence of a non-minimal derivative coupling. In such cases, it is anticipated that the violation of tensor CR occurs due to the violation of adiabaticity caused by the presence of a non-minimal derivative coupling. This violation does not occur in the conventional sense of a tensor mode evolving on superhorizon scales, but rather in a manner where the superhorizon mode cannot be regarded as a pure gauge mode due to its distinguishability in a local inertial frame. As a result, even if the superhorizon tensor mode appears frozen, it is not classified as an adiabatic mode, leading to an expected violation of tensor CRs. From a different perspective, this violation of tensor CR alongside a frozen tensor mode might be a distinctive signature of the presence of such non-minimal interactions with the tensor mode. This analysis can also be easily extended to other non-minimal derivative couplings. 
In general, a violation of the Maldacena CR might also indicate a violation of the EP so CRs also provide an interesting way to test the EP on large cosmological scales \cite{Creminelli:2013nua}. 

\section{Conclusions and discussions}
\label{conclusions}
In this paper, we have studied the cross-correlations of the inflationary scalar and tensor perturbations with the fluctuations of a non-minimally coupled spectator field with dilatonic coupling, providing valuable insights into correlation functions beyond the minimal setup. Firstly, we observed that during slow-roll inflation, the leading order interaction of the spectator and the scalar metric fluctuations in the comoving gauge originates from the dilatonic coupling, highlighting its significance, and found that the additional gravitational interaction from the non-minimal coupling is subject to slow-roll suppression. Notably, this fact became more apparent when considering the flat gauge. This observation led us to derive the CR for the scalar cross-correlation through a straightforward semi-classical approach. Importantly, this derivation represents a generalization of the CR for the cross-correlation of scalar metric fluctuations with gauge fields established in \cite{Jain:2012ga}. Our analysis demonstrated that the conformal nature of the spectator field is irrelevant to such semi-classical derivation, and it can be easily established in the flat gauge. In addition, these relations hold true in a generic manner, even in scenarios wherein the conventional semi-classical derivation, e.g., the Maldacena approach fails. We emphasise that these CRs have enormous potential which could be explored in various contexts. For instance, if we identify the spectator field as a potential isocurvature mode, then in cases where the isocurvature mode is directly coupled with the inflaton, these CRs become particularly valuable. They can also capture the NG associated with the isocurvature fluctuations within the inflationary context. The nature of these relations and their connection with non-linearly realized symmetries pose interesting questions for further exploration. Soft theorems (CRs) often arise as a result of such non-linear realizations, prompting an enticing avenue of investigation into the underlying symmetry and its non-linear manifestation within the framework of these novel CRs. We defer the pursuit of these interesting directions to our future work.

Further, we have also explored the cross-correlation of tensor perturbation with the spectator field and associated CRs. It is often observed that the tensor CRs are more robust relations and have remarkable universality compared to those for scalars. They remain valid even when there are multiple scalar fields as long as any anisotropies decrease rapidly in an exponential manner \cite{Bordin:2016ruc}. Contrary to the usual lore, our analysis shows that the violation of the tensor CR does not necessarily imply the existence of a non-freezing tensor mode. But this is not surprising because if we consider our working definition of an adiabatic mode as a specific superhorizon cosmological perturbation that is locally indistinguishable from a pure gauge mode, i.e., it can be absorbed by means of a suitable coordinate transformation. In the presence of a non-minimal derivative coupling, one can not treat the superhorizon mode as a pure gauge mode because it can be distinguished in a local inertial frame. Thus, even if the superhorizon tensor mode is frozen, it is not an adiabatic mode, and we expect a violation of the tensor CRs which can be considered a specific signature of the non-minimal derivative coupling.  

It is well known that light spectator fields usually induce isocurvature perturbations which might leave interesting imprints on cosmological observables \cite{Chung:2015pga}. 
In some specific scenarios such as the curvaton, they are converted to adiabatic perturbations at a later stage. The power spectrum of isocurvature modes is well constrained on large scales which are probed by CMB observations. However, they are not constrained on smaller scales. For instance, a scale-invariant or a very blue isocurvature perturbation spectrum may leave large effects on the short wavelength scales. Detection of such imprints will reveal the underlying high energy physics of the isocurvature
sector. Moreover, the NG associated with the isocurvature perturbations \cite{Linde:1996gt}, either in the form of a three-point correlation or a cross-correlation with curvature perturbations will also provide deeper insights into their generation mechanism. It might be interesting to explore if such isocurvature NG can help in the formation of primordial black holes on smaller scales. We leave these interesting directions for future work.

\section*{Acknowledgments}
We would especially like to thank Martin S. Sloth for suggesting the topic of this work, initial collaborations and numerous discussions. We also thank Chethan Krishnan for fruitful discussions.
RKJ wishes to acknowledge financial support from the new faculty seed start-up grant of the Indian Institute of Science, Bengaluru, India, Science and Engineering Research Board, Department of Science and Technology, Government of India, through the Core Research Grant~CRG/2018/002200, the MATRICS grant MTR/2022/000821 and the Infosys Foundation, Bengaluru, India through the Infosys Young Investigator award. PJS acknowledges Sarojini Damodaran Foundation for providing financial support for his visit to CP3-Origins, University of Southern Denmark.
\appendix
\section{Energy-momentum exchange relation for the spectator}
\label{EM_exchange}
In this section, we provide a detailed proof of eq. (\ref{energy_exchange}). The energy-momentum tensor $T_{\mu \nu}^{(\sigma)}$ corresponds to the spectator field $\sigma$ is obtained from action (\ref{action}) as,  
\bea
T_{\mu \nu}^{(\sigma)}=\lambda\left(\nabla_\mu\sigma \nabla_\nu \sigma-\frac{1}{2}g_{\mu \nu}\nabla_\alpha\sigma \nabla^\alpha \sigma\right)+\alpha \Theta^{(1)}_{\mu \nu}+\frac{\xi}{6} \Theta^{(2)}_{\mu \nu}~,
\eea
with, 
\bea 
\Theta^{(1)}_{\mu \nu}&=&\frac{1}{2}g_{\mu \nu}\left(\nabla_\alpha \nabla_\beta-R_{\alpha \beta}\right)\lambda\nabla^\alpha\sigma \nabla^\beta\sigma+\frac{1}{2}\Box\left( \lambda \nabla_\mu \sigma \nabla_\nu\sigma\right)+2 \lambda R_{\mu \alpha}\nabla^\alpha \sigma \nabla_\nu \sigma\nonumber\\&-&\frac{1}{2}\nabla^\alpha\nabla_\mu\left(\lambda \nabla_\nu \sigma \nabla_\alpha \sigma\right)-\frac{1}{2}\nabla^\alpha\nabla_\nu\left(\lambda \nabla_\mu \sigma \nabla_\alpha \sigma\right)~,\\
\Theta^{(2)}_{\mu \nu}&=&\left(G_{\mu \nu}+g_{\mu \nu}\Box-\nabla_\mu \nabla_\nu\right)\lambda \sigma^2~.
\eea 
In this work,  we only have to deal with the energy-momentum tensor of the spectator field. Therefore, we shall omit the superscript $(\sigma)$ and denote it with $T_{\mu \nu}$ for convenience.

One trivial way to prove eq. (\ref{energy_exchange}) is by explicitly working out the four divergences of the above-mentioned energy-momentum tensor using the equation of motion of $\sigma$. 
However, there exists a more general way of proving it without using the detailed form of the spectator Lagrangian. We just have to use the fact that it is directly coupled to the inflaton. For this purpose, let's assume the form of spectator action as, 
\bea 
\label{action_sigma}
S_\sigma=\int d^4x \sqrt{-g}\,\lambda(\phi) \mathcal{L}_\sigma~.
\eea 
Demanding the diffeomorphic invariance of this action gives us the desired result. To see that, let us consider an infinitesimal coordinate transformation from $x^\mu$ to $x^{\prime \mu}=x^\mu+\xi^\mu$. Under this transformation, one finds $\delta g^{\mu \nu}=\nabla^\mu \xi^\nu+\nabla^\nu\xi^\mu$ and the corresponding change in the action $S_\sigma$ can be written in the variational sense as, 
\bea 
\delta S_\sigma=\left(\frac{\delta S_\sigma}{\delta \phi} \right)_{g,\sigma}\delta\phi+\left(\frac{\delta S_\sigma}{\delta \sigma}\right)_{g,\phi} \delta\sigma+\left(\frac{\delta S_\sigma}{\delta g^{\mu \nu}}\right)_{\phi,\sigma}\delta g^{\mu \nu}~. 
\eea
Note that, the second term vanishes when the equation of motion of $\sigma$ is satisfied. For the first term, we find $\delta \phi=-\xi^\nu \partial_\nu\phi$, and thus 
\bea 
\frac{\delta S_\sigma}{\delta \phi} =\int d^4x\sqrt{-g}\, \frac{d\lambda}{d\phi}\, \mathcal{L}_\sigma \delta\phi =-\int d^4x\sqrt{-g}\,\mathcal{L}_\sigma(\nabla_\nu \lambda)\xi^\nu
\eea 
and the third term can be seen as 
\bea 
\frac{\delta S_\sigma}{\delta g^{\mu \nu}}\delta g^{\mu \nu}=-\frac{1}{2}\int d^4x \;\sqrt{-g} \;T_{\mu \nu}\delta g^{\mu \nu}=\int d^4x\sqrt{-g}\;(\nabla_\mu T^\mu_\nu) \xi^\nu-\int d^4x \sqrt{-g}\;\nabla_\mu(T^\mu_\nu\xi^\nu)~.
\eea 
The last term is a total divergent term and the natural boundary conditions on $\xi^\mu$ set it to be zero. Then, if we demand the diffeomorphic invariance of $S_\sigma$ left-hand side can be set to zero.
\bea 
\delta S_\sigma= \int d^4x\sqrt{-g} \bigg(\nabla_\mu T^\mu_\nu-(\nabla_\nu \lambda)\mathcal{L}_\sigma\bigg)\xi^\nu=0
\eea
This proves, 
\bea
\nabla_\mu T^{\mu\nu}=(\nabla^\nu \lambda)\mathcal{L}_\sigma~.
\eea 
This result can be used to easily derive the cubic order action as follows
\bea 
\label{cubic_action}
\delta S_{\zeta \sigma \sigma}=\frac{1}{2}\int d^4x \;\sqrt{-g} \;T^{\mu \nu}\delta g_{\mu \nu}=\int d^4x \sqrt{-g} \frac{\zeta}{H} \nabla_\mu T^{\mu 0}+\mathcal{O}(\epsilon)\simeq-\int d^4x \sqrt{-g} \dot{\lambda}\frac{\zeta}{H}\mathcal{L}_\sigma~.
\eea 
We again stress that the above results only applies to the action written in the form of eq. (\ref{action_sigma}) and does not depend on the explicit form of the Lagrangian.

\section{Evaluation of integrals in the squeezed limit}
\label{integrals}
In this section, we evaluate the integrals (\ref{I1}), (\ref{I2}) and (\ref{I3}) in squeezed limit, i.e., ${\bf k}_1 \to 0$. Thus throughout this section, we can approximate 
\bea 
\lim_{{\bf k}_1 \to 0}\zeta_{k_1}(\tau)=\frac{H}{\sqrt{4\epsilon k^3}}\approx \zeta_{k_1}(\tau_I) ~, 
\eea 
and we use eq. (\ref{modefn0}) as mode function $\sigma_{k}(\tau)$. However, all of the following analyses can also be trivially performed with eq. (\ref{Rij_modefn}). Utilizing these in all the integrals and considering the fact that in the squeezed limit $k_3\approx k_2=k$, we can express the integrals as follows:
\bea
\label{sqI1}
\mathcal{I}_1&=&-(4n)\; |\zeta_{k_1}(\tau_I)|^2\,{\rm Im} \left[\sigma^2_{k}(\tau_I)\int^{\tau_I} d\tau a^2 \lambda \left({\sigma'}_{k}^*(\tau)\right)^2\right]~, \\
\label{sqI2}
\mathcal{I}_2 &=&-(4n)\; |\zeta_{k_1}(\tau_I)|^2\,{\rm Im} \left[\sigma^2_{k}(\tau_I)\int^{\tau_I}  d\tau  a^2 \lambda \left({\sigma}_{k}^*(\tau)\right)^2\right]~,\\
\label{sqI3}
\mathcal{I}_3&=&-(4n)\;|\zeta_{k_1}(\tau_I)|^2\,{\rm Im} \left[\sigma^2_{k}(\tau_I)\int^{\tau_I}  \frac{d\tau}{\tau^2} a^2 \lambda \left({\sigma}_{k}^*(\tau)\right)^2\right]~. 
\eea
Here we have used $\lambda'/\lambda=-2n/\tau $, and $\tau_I$ denotes very late times at which we evaluate the correlator. 
In addition, one can also rewrite the eq. (\ref{eom1}) as follows, 
\bea 
\frac{d}{d\tau}\left(a^2\lambda\sigma_k'\right)+\left(k^2+a^2m^2+\xi \frac{a''}{a}\right)a^2\lambda\sigma_k=0~.
\eea 
This equation can now be used to rewrite the integral involved in eq. (\ref{sqI1}) as, 
\bea 
\int d\tau  a^2 \lambda \left({\sigma'}_{k}^*(\tau)\right)^2&=&a^2\lambda{\sigma}_{k}^*{\sigma'}_{k}^*-\int d\tau {\sigma}_{k}^*\frac{d}{d\tau}\left(a^2\lambda{\sigma'}_{k}^*\right)\nonumber \\
&=&a^2\lambda{\sigma}_{k}^*{\sigma'}_{k}^*+\int d\tau \;a^2 \lambda\left(k^2+\frac{2\xi+(m^2/H^2)}{\tau^2}\right)\left({\sigma}_{k}^*(\tau)\right)^2~.
\eea 
Upon using this result in eq. (\ref{sqI1}), we get
\bea 
\mathcal{I}_1=-(4n)\; |\zeta_{k_1}(\tau_I)|^2\,{\rm Im} \left[\sigma^2_{k}(\tau_I) \left[ a^2 \lambda \;{\sigma}_{k}^*(\tau){\sigma'}_{k}^*(\tau)\right]\bigg|^{\tau_I}_{-\infty}\right]+k^2\mathcal{I}_2+\left(2\xi+\frac{m^2}{H^2}\right)\mathcal{I}_3~.
\eea 
It can be easily shown that the lower limit will not contribute with in-in time contour, and this leads to the following relation
\bea 
\mathcal{I}_1=-(4n)\; |\zeta_{k_1}(\tau_I)|^2 |\sigma_{k}(\tau_I)|^2\,{\rm Im} \left[ a^2(\tau_I) \lambda(\tau_I) \;\sigma_k(\tau_I){\sigma'}_{k}^*(\tau_I)\right]+k^2\mathcal{I}_2+\left(2\xi+\frac{m^2}{H^2}\right)\mathcal{I}_3~.
\eea 
The Wronskian corresponds to the equation of motion (\ref{eom1}) and the Bunch-Davis initial condition determines, ${\rm Im} \left[\sigma_k(\tau_I){\sigma'}_{k}^*(\tau_I)\right]=1/(2a^2_I\lambda_I )$. Here the subscript '$I$' denotes the corresponding quantities evaluated at time $\tau=\tau_I$. Then we obtain, 
\bea 
\label{integral-relation}
\mathcal{I}_1=-(2n)\; |\zeta_{k_1}(\tau_I)|^2 |\sigma_{k}(\tau_I)|^2+k^2\mathcal{I}_2+\left(2\xi+\frac{m^2}{H^2}\right)\mathcal{I}_3~.
\eea
This result is sufficient to evaluate the squeezed limit of the correlator given in eq. (\ref{FIN}). Upon utilizing this result in the squeezed limit of (\ref{FIN}), one can see that the integrals $\mathcal{I}_2$ and $\mathcal{I}_3$ nicely cancel out. So we do not need to evaluate these integrals explicitly to compute the squeezed limit of the correlator. 
However, let's demonstrate the evaluation of $\mathcal{I}_2$ below because the same procedure can be used to evaluate the integral involved in section \ref{tensor-corr}. Such integrals can be evaluated using the same method as previously outlined in \cite{Jain:2012vm,Jain:2021pve}. Thus, let us express the integral $\mathcal{I}_2$ in the squeezed limit as, 
\bea 
\label{app_I2}
\mathcal{I}_2 =-2|\zeta_{k_1}(\tau_I)|^2|\sigma_{k_2}(\tau_I)|^2\,{\rm Im} \left[e^{i(\nu+1/2)\pi}\int d\tau \tau a^2 \lambda'(\tau) \left(\sigma_{k_2}^*(\tau)\right)^2\right]~,
\eea 
where we have used $\sigma^2_{k}(\tau_I)\approx -|\sigma_{k}(\tau_I)|^2e^{i\pi(\nu+1/2)}$ in above equation. Now, using the explicit form of the mode function and 
\bea
\label{Idenity_H2}
{\rm Im}\left[\int_0^{\infty}dx x \left(H^{(2)}_{\nu}(x)\right)^2\right]=\frac{2\nu}{\pi}~, 
\eea
we obtain
\bea 
\mathcal{I}_2= \left(\frac{2n\nu}{k^2_2}\right)|\zeta_{k_1}(\tau_I)|^2|\sigma_{k_2}(\tau_I)|^2~.
\eea
\bibliographystyle{JHEP} 
\bibliography{ref}
\end{document}